# Vectorial Acoustic Multiplexed Holography

Yuan Tian[a], Hao Ge[a,b], Jiangpo Zheng[a], Xiujuan Zhang[a], Ming-Hui Lu[a, b, c, d *], and Yan-Feng Chen[a, d *]

[a]*State Key Laboratory of Solid State Microstructures and Department of Materials Science and Engineering, Nanjing University, Nanjing 210093, China*

[b]*School of Advanced Manufacturing Engineering, Nanjing University, Suzhou, 215163, China*

[c]*Jiangsu Key Laboratory of Artificial Functional Materials, Nanjing 210093, China*

[d]*Collaborative Innovation Center of Advanced Microstructures, Nanjing University, Nanjing 210093, China*

[*]Corresponding author: luminghui@nju.edu.cn (Ming-Hui Lu); yfchen@nju.edu.cn (Yan-Feng Chen)

## Abstract

Encoding more information into wave fields is a central goal in imaging, communication, and wave control. Optical holography benefits from polarization multiplexing, but acoustic holography remains largely limited to pressure-only encoding because sound in fluids lacks naturally independent vector channels. Here, we show that particle velocity can serve as a practical multiplexing degree of freedom despite the intrinsic pressure-velocity coupling governed by the acoustic Euler equation. We develop a physics-informed inverse-design approach that incorporates acoustic propagation and pressure-velocity coupling to create a binary metasurface for vector-field acoustic holographic multiplexing. Experiments demonstrate dual-channel multiplexing on the in-plane velocity components $v_x$ and $v_y$, and further extend to three-channel multiplexing by incorporating pressure $p$, with high-fidelity reconstruction and low cross-talk. This approach adds a new information dimension without reducing spatial or spectral bandwidth and enables broader forms of wave-based information encoding and multiplexed wave control.

# 1. Introduction

Holography reconstructs complex wavefields with high spatial fidelity and supports applications such as volumetric displays, high-density data storage, information encryption, and contactless manipulation [1–9]. A central goal in holography is to increase information capacity, namely, the number of independent channels that can be encoded and retrieved in parallel. This capacity has been expanded by multiplexing spatial, spectral, and orbital angular momentum degrees of freedom [10–16]. In optics, it is further enhanced by polarization multiplexing, where orthogonal polarization states provide naturally independent channels [17–19].

Acoustic holography offers distinct advantages over light, including deeper penetration in opaque media and stronger radiation forces for manipulation, which support applications from medical ultrasound to acoustic tweezing [20–23]. However, sound waves in fluids are longitudinal rather than transverse, and therefore do not provide a polarization degree of freedom [24]. As a result, acoustic holography is usually developed in a pressure-only framework, and multiplexing has mainly relied on spatial or spectral dimensions [25]. Although particle motion in acoustic fields is vectorial, this vector character has not yet provided practical holographic channels comparable to optical polarization. This raises a key question: can usable vector channels be created in acoustics despite the absence of naturally independent polarization states?

Particle velocity offers a possible route to such channels. Unlike acoustic pressure, which is scalar, particle velocity is a vector field and is linked to rich wave phenomena such as acoustic spin and skyrmions [26–29]. This makes it natural to ask whether different particle velocity components can be used to encode different holographic images. This challenge, however, is not simply to generate multiple outputs. The particle velocity components are not free variables. In fluids, they are determined by spatial gradients of the same scalar pressure field through Euler's relation [24]. Multiple channels therefore originate from a shared scalar field, and shaping one component generally alters the others. Without the natural orthogonality available in optical polarization, independent multi-channel encoding becomes a strongly constrained inverse problem. Recent progress in metasurface holography [30–37] and physics-based deep learning [38–43] has created new opportunities for solving this type of coupled inverse-design problem.

Here, we introduce a vector-field acoustic holography multiplexing strategy that uses particle velocity components as additional information channels. By incorporating the physical coupling between pressure and velocity into the inverse-design process, we design a single metasurface that forms functionally independent holographic images on coupled in-plane velocity components. We experimentally demonstrate dual-channel multiplexing on $v_x$ and $v_y$, and further extend the scheme to three-channel multiplexing by incorporating pressure $p$, with low cross-talk. These results show that coupled particle velocity components can serve as practical multiplexing channels for acoustic holography, adding a new information dimension without reducing spatial or spectral bandwidth.

**2. Materials and Methods**

*2.1. Physics-informed inverse design framework*

We employ a physics-informed deep learning framework based on a U-net architecture for the inverse design. The network takes normalized target images (corresponding to $v_x$, $v_y$, and optionally $p$) as input and predicts a single-channel amplitude distribution for the metasurface. To ensure manufacturability, a Gumbel-sigmoid activation is applied to constrain the output to binary states {0, 1}, enabling hard binarization in the forward pass while maintaining gradient flow during back-propagation.

The network is trained end-to-end by minimizing the MSE between the target and reconstructed fields. Crucially, the optimization is guided by a differentiable acoustic forward model that integrates the angular spectrum method (ASM) with Euler's relation, ensuring that the intrinsic physical coupling is embedded into the learning process. Training is performed using the Adam optimizer (learning rate $10^{-3}$) over 2000 iterations in PyTorch. This framework is highly versatile; beyond binary amplitude modulation, it is readily adaptable to phase-only modulation by simply removing the binarization constraint to directly optimize continuous phase profiles (see Supplementary Note 4).

*2.2. Numerical simulation*

Numerical simulations are performed in COMSOL Multiphysics using the *Pressure Acoustics, Boundary Element* module in the frequency domain. This full-wave approach solves the Helmholtz equation in its boundary integral form, enabling accurate modeling of acoustic wave propagation in open domains without artificial boundaries. By prescribing the amplitude and phase distribution on the metasurface plane, the resulting pressure and particle velocity fields in the image plane are computed and used to validate the designed holograms.

*2.3. Experiments*

Binary metasurface samples are fabricated by etching 304 stainless-steel plates with a machining accuracy of 0.01 mm. An ultrasonic transducer with a diameter of 80 mm is used as the sound source and positioned 1.2 m in front of the metasurface; the detailed experimental setup and the characterization of the incident acoustic field are provided in Supplementary Note 1 and Fig. S1. The transducer is driven by a signal generator through a power amplifier, while a reference signal is simultaneously fed into a data acquisition system (National Instruments PCI-6251) for phase retrieval. The acoustic field is measured using a 0.25-inch microphone (Brüel & Kjær 4939) mounted on a three-axis translation stage to scan the holographic image plane with a step size of 2.34 mm. The data acquisition system synchronously records both the microphone and reference signals to reconstruct the complex pressure field (amplitude and phase). Finally, the particle velocity field is derived from the reconstructed pressure field via numerical differentiation based on Euler's relation.

## 3. Results and discussion

### 3.1. Principle of vector-field acoustic multiplexing

Figure 1 schematically illustrates the concept of vector-field acoustic holographic multiplexing. In our strategy, the in-plane particle velocity components $v_x$ and $v_y$ are used as parallel holographic channels (Fig. 1a), and the scheme can be further extended by incorporating the pressure field $p$ as an additional channel (Fig. 1b).

The propagation of sound waves in fluids is governed by the linearized Euler equation,

$$\rho_0 \frac{\partial v}{\partial t} = -\nabla p, \tag{1}$$

where $v$ denotes the particle velocity, $p$ is the acoustic pressure, and $\rho_0$ is the ambient density. For time-harmonic waves with temporal dependence $e^{-j\omega t}$, Eq. (1) becomes

$$\boldsymbol{v} = -\frac{j}{\rho \omega} \nabla p. \tag{2}$$

Equation (2) shows that the particle velocity components are determined by spatial gradients of the same scalar pressure field. Unlike orthogonal polarization states in optics, $v_x$ and $v_y$ are therefore physically coupled, which makes independent multi-channel image encoding a strongly constrained inverse problem.

To address this challenge, we do not seek strict physical orthogonality. Instead, we aim for functional independence at the image level. Using a machine-learning-assisted inverse-design framework, we optimize the scalar pressure field so that its spatial gradients simultaneously produce the target intensity distributions in the $v_x$ and $v_y$ channels. In this way, the coupled velocity components can be used as practical holographic channels for vector-field acoustic multiplexing.

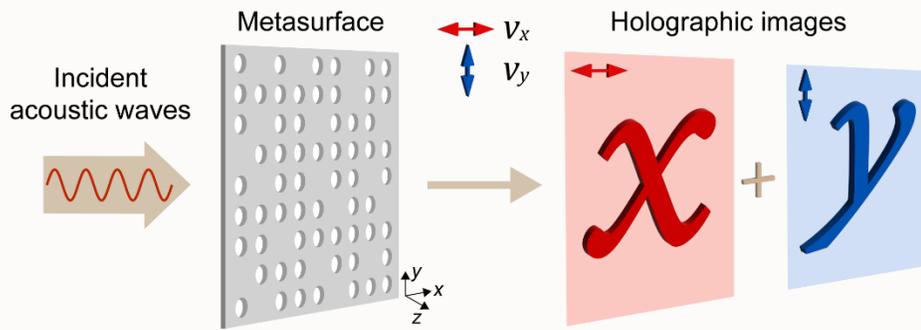

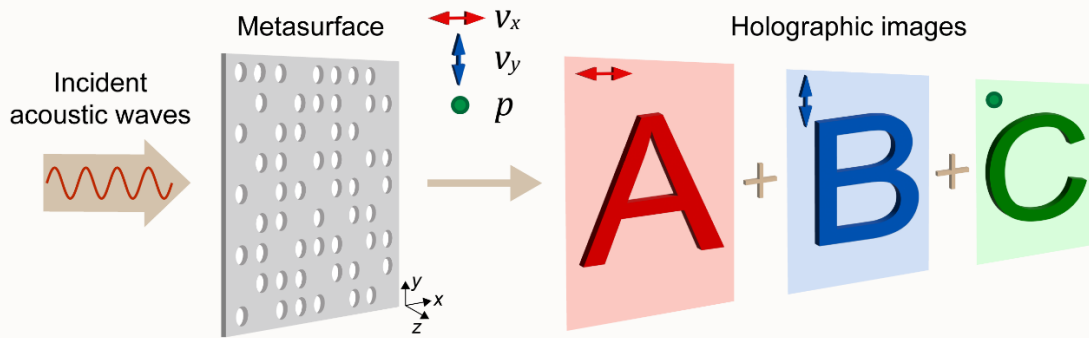

**Fig. 1.** Principle of vector-field acoustic holographic multiplexing. (a) Schematic illustration of the multiplexing strategy. An incident acoustic wave is modulated by a metasurface, and the transmitted field is decomposed into its in-plane velocity components $v_x$ and $v_y$, which serve as independent holographic channels to reconstruct distinct target images in parallel. (b) Extension of the concept by including the scalar pressure field $p$ as an additional channel, enabling three-channel holography with enhanced information capacity.

*3.2. Physics-informed inverse design framework*

To realize vector-field acoustic holographic multiplexing, we use an end-to-end inverse-design framework that combines a neural network with a differentiable acoustic forward model (Fig. 2a). Unlike conventional black-box models, this framework explicitly incorporates the governing physics into the optimization. It takes the target images as input and iteratively updates the metasurface structure until the reconstructed vector fields converge to the desired holographic outputs.

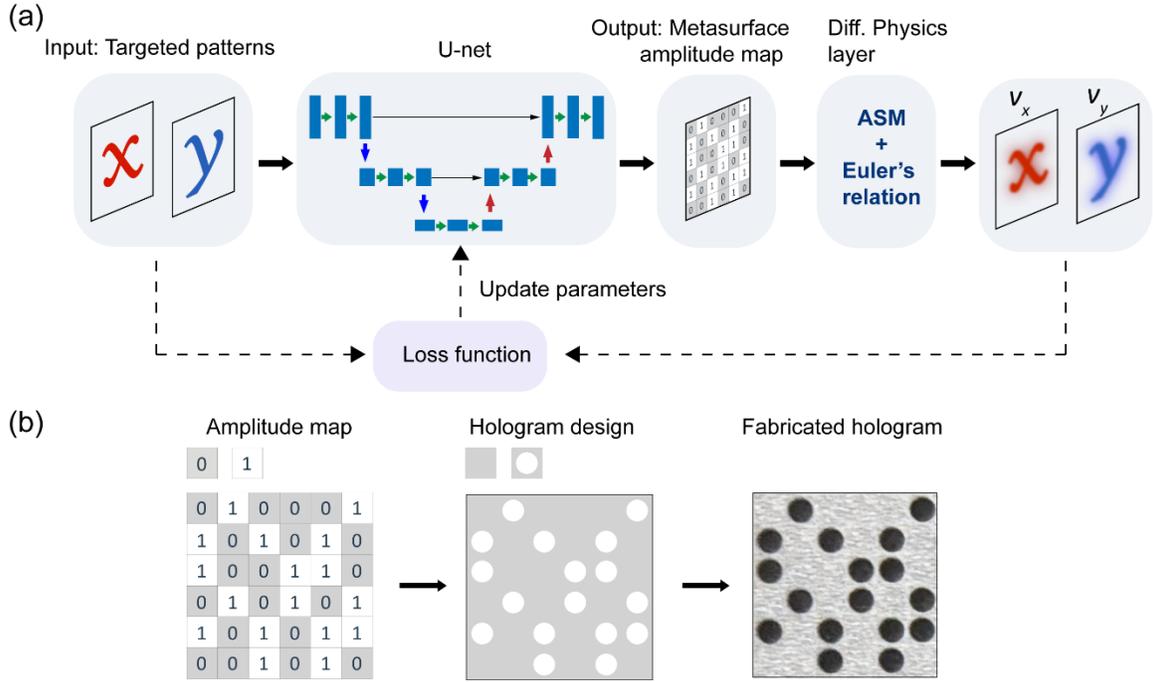

**Fig. 2.** Physics-informed inverse design of binary acoustic metasurfaces. (a) Machine-learning-assisted framework for vector-field holography. Crucially, the network output is passed through a non-trainable differentiable physics layer. This layer explicitly implements the angular spectrum method and Euler's relation, serving as a gradient-permeable acoustic propagator that physically maps the metasurface design to the coupled vector velocity fields. The discrepancy between reconstructed and target images defines a multi-channel loss function. (b) Conversion of optimized amplitude maps into binary (0/1) transmission patterns, which can be directly fabricated while retaining effective control of vector-field holograms.

The core engine is a U-net–based convolutional neural network featuring a symmetric encoder–decoder architecture. The encoder extracts multi-scale hierarchical features through successive convolution and pooling operations, while the decoder restores spatial resolution via transposed convolutions and skip connections. For dual-channel designs, the network inputs are two normalized target images corresponding to the particle velocity components $v_x$ and $v_y$; for the three-channel design, the pressure-field target $p$ is included as an additional input. The network outputs a single-channel map representing the metasurface transmission function $T_{\text{meta}}(x, y)$, whose values are constrained to the binary set {0, 1} to facilitate fabrication.

Crucially, the optimization pipeline includes a dedicated "differentiable physics layer" following the network output. Unlike standard black-box training, the predicted metasurface transmission function is fed directly into this non-trainable layer, which rigorously solves the wave propagation equation using the ASM. Given an incident pressure field $p_{\text{in}}(x,y)$ and the metasurface transmission function $T_{\text{meta}}(x,y)$, the transmitted pressure at the metasurface plane is

$$p_0(x,y) = p_{\text{in}}(x,y) \cdot T_{\text{meta}}(x,y). \tag{3}$$

The corresponding angular spectrum propagates to a distance $z$ according to

$$P(k_x, k_y, z) = P_0(k_x, k_y) e^{jz\sqrt{k^2 - k_x^2 - k_y^2}}, \tag{4}$$

and the pressure distribution at the target plane is obtained through the inverse Fourier transform

$$p(x,y,z) = \mathcal{F}^{-1}\{P(k_x, k_y, z)\}. \tag{5}$$

The particle velocity components follow directly from Euler's relation:

$$v_x = -\frac{j}{\rho_0 \omega} \frac{\partial p}{\partial x}, \tag{6}$$

$$v_y = -\frac{j}{\rho_0 \omega} \frac{\partial p}{\partial y}. \tag{7}$$

By embedding these governing equations (Eqs. 3-7) into a gradient-permeable computational graph, the error gradients computed from the vector fields are back-propagated through the Euler coupling and the diffraction operators directly to the metasurface parameters. This guides the network toward solutions that satisfy the physical constraints while producing functionally independent image channels.

During training, the discrepancy between the reconstructed and target intensity distributions drives the optimization via a multi-channel loss function. For dual-channel multiplexing, the loss is defined as the mean-squared error (MSE) averaged across the velocity channels:

$$Loss = \frac{1}{K \cdot N} \sum_{i \in \{v_x, v_y\}} \left\| I_i^{\text{pred}} - I_i^{\text{target}} \right\|^2, \tag{8}$$

where $I_i^{\text{pred}}$ and $I_i^{\text{target}}$ denote the predicted and target intensity distributions of channel $i$, $N$ represents the total number of pixels in a single image, $K$ is the number of multiplexed channels. For the three-channel system, the scalar pressure channel $p$ is incorporated analogously. Detailed analysis of the optimization convergence and the visual evolution of the vector fields during training are presented in Supplementary Note 2.

To ensure manufacturability and robustness, we employ a Gumbel-sigmoid activation to constrain the metasurface design to a binary {0, 1} amplitude pattern. Such a binary structure significantly simplifies fabrication and minimizes accumulated phase errors. The optimized amplitude map is directly converted into the perforated metasurface layout shown in Fig. 2B for experimental realization. This framework is highly versatile; beyond binary amplitude modulation, it is readily adaptable to phase-only modulation by simply removing the binarization constraint to directly optimize continuous phase profiles (see Supplementary Note 4).

*3.3. Dual-channel vector-field holography*

To experimentally validate the principle of vector-field holographic multiplexing, we first constructed a dual-channel system in which the two orthogonal in-plane components of the particle velocity field, $v_x$ and $v_y$, serve as independent holographic channels. As illustrated in Fig. 3a, the letters "X" and "Y" are chosen as the target images of the two channels, respectively. Based on these targets, the trained neural network generates a binary amplitude distribution (Fig. 3b), which is used to fabricate a metasurface hologram (Fig. 3c). The sample consists of a $300 \times 300$ mm² stainless-steel plate (corresponding to $20\lambda \times 20\lambda$ at 22.9 kHz, where $\lambda \approx 15$ mm), discretized into $128 \times 128$ subwavelength pixels with a pixel size of approximately $0.16\lambda$. The metasurface thickness is only 0.5 mm ($\sim\lambda/30$), which is negligible in the propagation model.

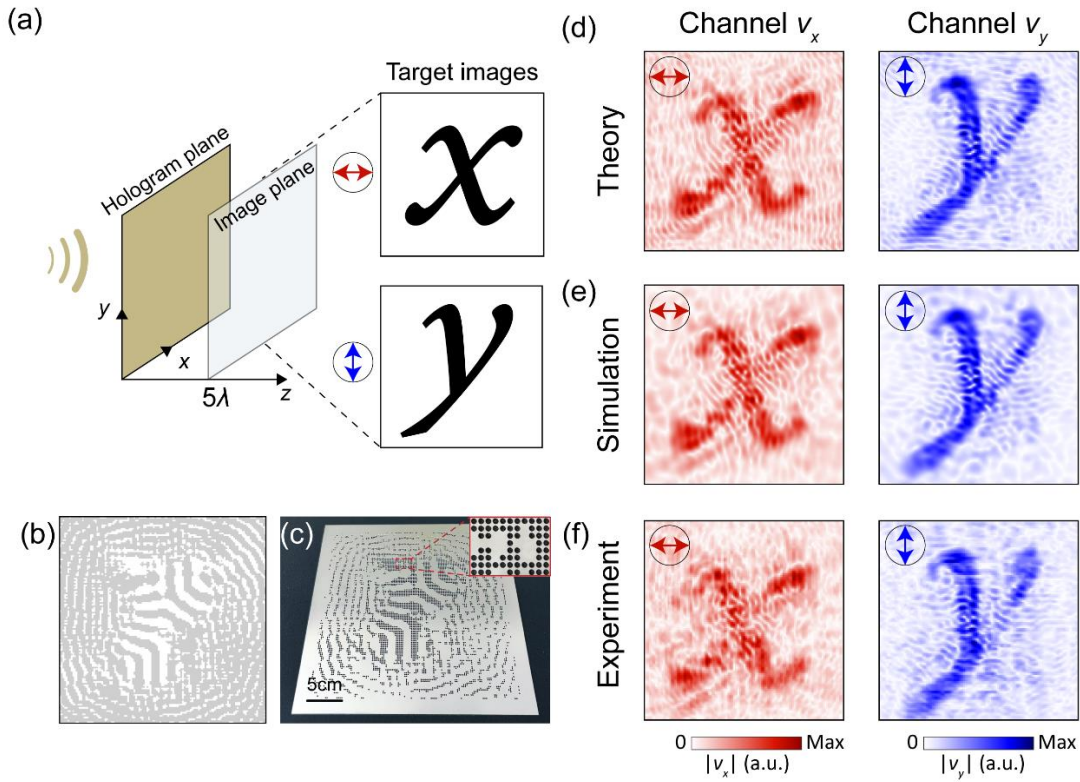

**Fig. 3.** Dual-channel vector-field holography. (a) Target images "X" and "Y" assigned to the $v_x$ and $v_y$ channels. (b) Binary amplitude distribution generated by the neural network for the metasurface design. (c) Photograph of the fabricated binary metasurface hologram. (d-f) Reconstructed images at the target plane from theory (d), simulation (e), and experiments (f), showing clear recovery of "X" in the $v_x$ channel and "Y" in the $v_y$ channel with minimal cross-talk.

We evaluate the holographic performance at a reconstruction plane located at $z = 5\lambda$ through theoretical calculations, full-wave simulations, and experimental measurements (Figs. 3d-f; see Methods for implementation details). The theoretical, simulated, and experimental results agree well: the "X" image is clearly reconstructed in the $v_x$ channel, while the "Y" pattern emerges in the $v_y$ channel with minimal interference, confirming the effectiveness of dual-channel vector-field multiplexing.

To quantitatively assess image fidelity, we compute the Pearson correlation coefficient, $C_i$, between the target ($T_i$) and reconstructed ($R_i$) intensity distributions:

$$C_i = \text{corr}(T_i, R_i) = \frac{\text{COV}(T_i, R_i)}{\sqrt{D(T_i) \cdot D(R_i)}}, \tag{9}$$

where COV denotes covariance and $D$ denotes variance. The correlation values for the $v_x$ and $v_y$ channels are 0.80 / 0.80 / 0.63 and 0.80 / 0.83 / 0.72 for theory, simulation, and experiment, respectively. All values exceed 0.6, confirming high-fidelity image formation.

To quantify the channel isolation, we introduce a differential cross-talk index, $X_{i,j}$, which calibrates the inter-channel correlation against the intrinsic similarity of the target patterns:

$$X_{i,j} = |\text{corr}(R_i, R_j) - \text{corr}(T_i, T_j)|, \tag{10}$$

where $\text{corr}(\cdot,\cdot)$ denotes the correlation coefficient defined in Eq. (9). This metric decouples the structural similarity inherent to the target images from system-induced interference, thereby isolating the net information leakage introduced by the multiplexing mechanism. The calculated differential cross-talk indices between the two velocity channels are 0.15 / 0.20 / 0.15 for theory, simulation, and experiment, respectively. These low values ($\leq 0.2$) indicate weak interference between the two reconstructed channels. To further demonstrate the universality of our method regarding spatially complex patterns, we also successfully reconstructed intricate targets (e.g., fruits and doves), as detailed in Supplementary Note 3.

*3.4. Three-channel vector–scalar holography*

To extend vector-field multiplexing beyond two channels, we further introduce the scalar pressure field $p$ as a third independent channel, forming a three-channel vector–scalar holographic system. In this configuration, the letters "A", "B" and "C" are assigned to the $v_x$, $v_y$ and $p$ channels, respectively, representing three distinct pieces of encoded information (Fig. 4a). The target images are jointly optimized by the neural network, yielding a binary amplitude distribution (Fig. 4b) that is fabricated into a metasurface hologram (Fig. 4c) under identical geometric and experimental conditions to the dual-channel sample.

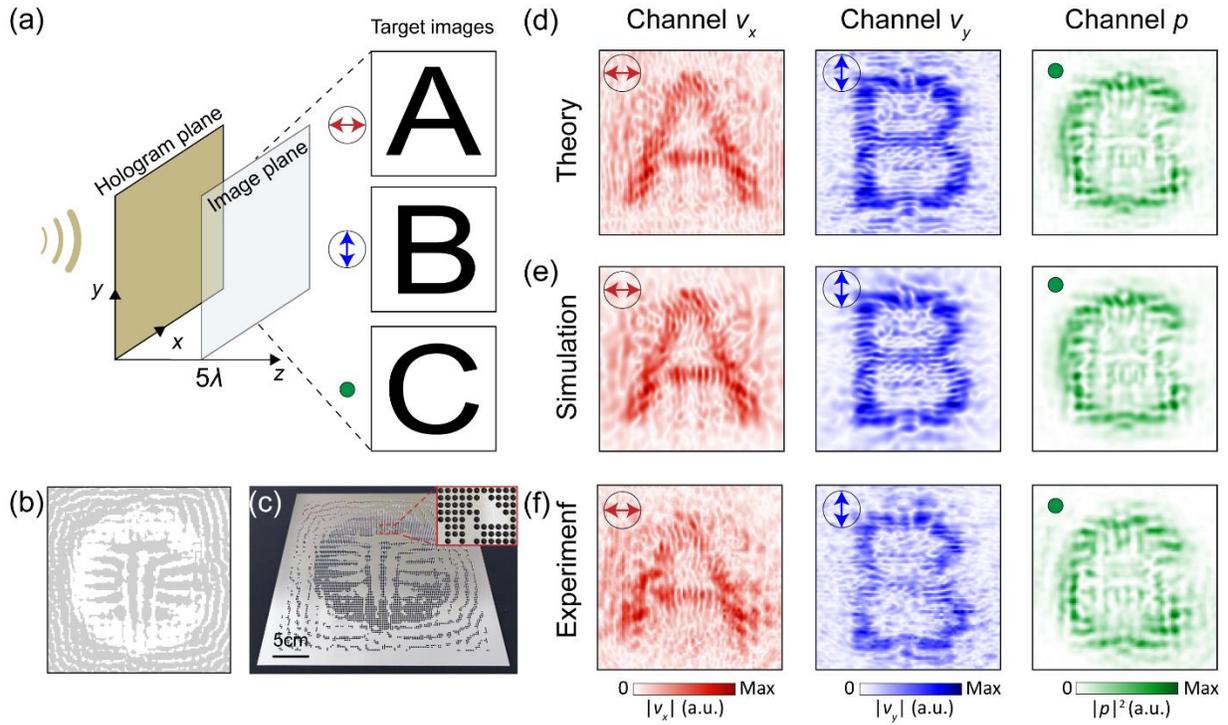

**Fig. 4.** Three-channel vector–scalar holography. (a) Target images "A", "B", and "C" assigned to the $v_x$, $v_y$ and pressure channels, respectively. (b) Binary amplitude distribution obtained from the joint optimization of the three channels. (c) Photograph of the fabricated metasurface hologram. (d–f) Reconstructed images from theory (d), simulations (e), and experiments (f), showing accurate recovery of "A" in the $v_x$ channel, "B" in the $v_y$ channel, and "C" in the pressure channel with negligible cross-talk.

The theoretical, simulated, and experimental reconstructions (Figs. 4d-f) show good agreement. The images of "A", "B", and "C" are clearly resolved in the $v_x$, $v_y$, and $p$ channels, respectively. Quantitatively, the reconstruction fidelity—characterized by the correlation coefficients—reaches 0.74 / 0.76 / 0.75 for $v_x$, 0.76 / 0.79 / 0.76 for $v_y$, and 0.75 / 0.76 / 0.56 for the pressure channel (for theory, simulation, and experiment, respectively). All theoretical and simulated values exceed 0.74, and experimental values remain robustly above 0.55, demonstrating effective information retrieval despite the complexity of three-channel multiplexing.

Channel independence is rigorously assessed using the differential cross-talk index ($X_{i,j}$) defined in Eq. (10). The calculated indices for the channel pairs ($v_x, v_y$), ($v_y, p$), and ($v_x, p$) are 0.03 / 0.09 / 0.01, 0.003 / 0.05 / 0.03, and 0.27 / 0.32 / 0.31 (for theory, simulation, and experiment), respectively. While a moderate residual coupling (indices ≈ 0.3) is observed between $v_x$ and $p$, the majority of channel pairs exhibit negligible interference with indices below 0.1. Importantly, this residual coupling does not compromise the visual distinguishability of the reconstructed patterns, confirming that latent vector degrees of freedom can be activated and multiplexed alongside the scalar pressure field. This result shows that vector and scalar channels can be combined within the same physical footprint and bandwidth to increase information capacity.

## 4. Conclusion

In this work, we experimentally demonstrate an acoustic holographic multiplexing strategy that uses particle velocity components as additional image channels. Using a physics-informed inverse design framework that accounts for the coupling between pressure and velocity, we realize both dual-channel and three-channel vector-scalar multiplexing. The experimental results show high-fidelity image reconstruction in each channel with low cross-talk, demonstrating that particle velocity components can serve as practical multiplexing channels in acoustic holography.

Compared with conventional pressure-only holography, this vector-field approach introduces additional degrees of freedom for acoustic field control and increases information capacity within the same physical footprint and bandwidth. It can also be combined with frequency and spatial multiplexing to further increase the information capacity of acoustic holography, as shown in Supplementary Note 5. Although our proof-of-concept experiments were performed in air, the same physical mechanism is relevant to underwater acoustics and ultrasound, where efficient use of available channels is important. Looking ahead, combining this mechanism with programmable or active acoustic materials may enable reconfigurable acoustic vector-field control.

**Conflict of interest**

The authors declare that they have no conflict of interest.

**Acknowledgments**

We acknowledge support from the National Key R&D Program of China (Grants No.

2023YFA1406904 and No. 2023YFA1407700), the National Natural Science Foundation of China (Grant No.52250363 and No. 12222407) and the Key R&D Program of Jiangsu Province (Grant No. BK20232015 and No. BK20232048). Y.T. thanks the supports from the China Postdoctoral Science Foundation (Grant No. 2023M731614 and No. 2023T160298), the Jiangsu Funding Program for Excellent Postdoctoral Talent (Grants No. 2023ZB114).

**Author contributions**

Y.T., M.-H.L., and Y.-F.C. conceived the original idea and supervised the project. Y.T. carried out the theoretical calculations, numerical simulations, and experimental measurements. Y.T. and M.-H.L. performed the data analysis. Y.T. and M.-H.L. wrote the manuscript. All authors contributed to the scientific discussion of the work.

**Appendix A. Supplementary materials**

Supplementary materials to this article can be found online at xxxxx.

**References**

[1] Tay S, Blanche P-A, Voorakaranam R, Tunç A V., Lin W, Rokutanda S, et al. An updatable holographic three-dimensional display. Nature 2008;451:694–8.

[2] Blanche P-A, Bablumian A, Voorakaranam R, Christenson C, Lin W, Gu T, et al. Holographic three-dimensional telepresence using large-area photorefractive polymer. Nature 2010;468:80–3.

[3] Smalley DE, Nygaard E, Squire K, Van Wagoner J, Rasmussen J, Gneiting S, et al. A photophoretic-trap volumetric display. Nature 2018;553:486–90.

[4] Heanue JF, Bashaw MC, Hesselink L. Volume holographic storage and retrieval of digital data. Science 1994;265:749–52.

[5] Hesselink L, Orlov SS, Liu A, Akella A, Lande D, Neurgaonkar RR. Photorefractive Materials for Nonvolatile Volume Holographic Data Storage. Science 1998;282:1089–94.

[6] Li J, Kamin S, Zheng G, Neubrech F, Zhang S, Liu N. Addressable metasurfaces for dynamic holography and optical information encryption. Sci Adv 2018;4:1–7.

[7] Qu G, Yang W, Song Q, Liu Y, Qiu C-W, Han J, et al. Reprogrammable meta-hologram for optical encryption. Nat Commun 2020;11:5484.

[8] Grier DG. A revolution in optical manipulation. Nature 2003;424:810–6.

[9]   Huft PR, Kolbow JD, Thweatt JT, Lindquist NC. Holographic Plasmonic Nanotweezers for Dynamic Trapping and Manipulation. Nano Lett 2017;17:7920–5.

[10]  Shi J, Qiao W, Hua J, Li R, Chen L. Spatial multiplexing holographic combiner for glasses-free augmented reality. Nanophotonics 2020;9:3003–10.

[11]  Li X, Ren H, Chen X, Liu J, Li Q, Li C, et al. Athermally photoreduced graphene oxides for three-dimensional holographic images. Nat Commun 2015;6:6984.

[12]  Li X, Chen L, Li Y, Zhang X, Pu M, Zhao Z, et al. Multicolor 3D meta-holography by broadband plasmonic modulation. Sci Adv 2016;2:1–6.

[13]  Lim KTP, Liu H, Liu Y, Yang JKW. Holographic colour prints for enhanced optical security by combined phase and amplitude control. Nat Commun 2019;10:1–8.

[14]  Ren H, Briere G, Fang X, Ni P, Sawant R, Héron S, et al. Metasurface orbital angular momentum holography. Nat Commun 2019;10:2986.

[15]  Fang X, Ren H, Gu M. Orbital angular momentum holography for high-security encryption. Nat Photonics 2020;14:102–8.

[16]  Shi Z, Wan Z, Zhan Z, Liu K, Liu Q, Fu X. Super-resolution orbital angular momentum holography. Nat Commun 2023;14:1–13.

[17]  Zhao R, Sain B, Wei Q, Tang C, Li X, Weiss T, et al. Multichannel vectorial holographic display and encryption. Light Sci Appl 2018;7:95.

[18]  Xiong B, Liu Y, Xu Y, Deng L, Chen C-W, Wang J-N, et al. Breaking the limitation of polarization multiplexing in optical metasurfaces with engineered noise. Science 2023;379:294–9.

[19]  Wang J, Chen J, Yu F, Chen R, Wang J, Zhao Z, et al. Unlocking ultra-high holographic information capacity through nonorthogonal polarization multiplexing. Nat Commun 2024;15:1–10.

[20]  Cummer SA, Christensen J, Alù A. Controlling sound with acoustic metamaterials. Nat Rev Mater 2016;1:16001.

[21]  Marzo A, Seah SA, Drinkwater BW, Sahoo DR, Long B, Subramanian S. Holographic acoustic elements for manipulation of levitated objects. Nat Commun 2015;6:8661.

[22]  Derayatifar M, Habibi M, Bhat R, Packirisamy M. Holographic direct sound printing. Nat Commun 2024;15:6691.

[23]  Ma Z, Holle AW, Melde K, Qiu T, Poeppel K, Kadiri VM, et al. Acoustic Holographic Cell Patterning in a Biocompatible Hydrogel. Adv Mater 2020;32:1–6.

[24]  Morse PM, Ingard KU. Theoretical acoustics. Princeton university press; 1986.

[25]  Melde K, Mark AG, Qiu T, Fischer P. Holograms for acoustics. Nature 2016;537:518–22.

[26] Shi C, Zhao R, Long Y, Yang S, Wang Y, Chen H, et al. Observation of acoustic spin. Natl Sci Rev 2019;6:707–12.

[27] Long Y, Zhang D, Yang C, Ge J, Chen H, Ren J. Realization of acoustic spin transport in metasurface waveguides. Nat Commun 2020;11:4716.

[28] Ge H, Xu XY, Liu L, Xu R, Lin ZK, Yu SY, et al. Observation of Acoustic Skyrmions. Phys Rev Lett 2021;127:144502.

[29] Hu P, Wu HW, Sun WJ, Zhou N, Chen X, Yang YQ, et al. Observation of localized acoustic skyrmions. Appl Phys Lett 2023;122.

[30] Zhang J, Tian Y, Cheng Y, Liu X. Acoustic holography using composite metasurfaces. Appl Phys Lett 2020;116.

[31] Assouar B, Liang B, Wu Y, Li Y, Cheng J-C, Jing Y. Acoustic metasurfaces. Nat Rev Mater 2018;3:460–72.

[32] Xie B, Tang K, Cheng H, Liu Z, Chen S, Tian J. Coding Acoustic Metasurfaces. Adv Mater 2017;29:1–8.

[33] Díaz-Rubio A, Tretyakov SA. Acoustic metasurfaces for scattering-free anomalous reflection and refraction. Phys Rev B 2017;96:1–10.

[34] Xie Y, Wang W, Chen H, Konneker A, Popa B-I, Cummer SA. Wavefront modulation and subwavelength diffractive acoustics with an acoustic metasurface. Nat Commun 2014;5:5553.

[35] Chen J, Xiao J, Lisevych D, Shakouri A, Fan Z. Deep-subwavelength control of acoustic waves in an ultra-compact metasurface lens. Nat Commun 2018;9:4920.

[36] Zhang M, Jin B, Hua Y, Zhu Z, Xu D, Fan Z, et al. Reconfigurable dynamic acoustic holography with acoustically transparent and programmable metamaterial. Nat Commun 2025;16:9126.

[37] Zhou Y, Li F, Wu K, Wang W, Liu J, Liang B, et al. Orbital angular momentum- and frequency-dependent high-capacity encrypted hologram through multi-dimensional multiplexing acoustic metasurface. Nat Commun 2025;16:11692.

[38] Zhong C, Sun Z, Li J, Jiang Y, Su H, Liu S. Accurate and real-time acoustic holography using super-resolution and physics combined deep learning. Appl Phys Lett 2025;126.

[39] Sui X, He Z, Chu D, Cao L. Non-convex optimization for inverse problem solving in computer-generated holography. Light Sci Appl 2024;13:158.

[40] Gopakumar M, Lee G-Y, Choi S, Chao B, Peng Y, Kim J, et al. Full-colour 3D holographic augmented-reality displays with metasurface waveguides. Nature 2024;629:791–7.

[41] Lin Q, Zhang R, Cai F, Chen Y, Ye J, Wang J, et al. Multi-frequency acoustic hologram generation with a physics-enhanced deep neural network. Ultrasonics 2023;132:106970.

[42] Shimobaba T, Blinder D, Birnbaum T, Hoshi I, Shiomi H, Schelkens P, et al. Deep-Learning Computational Holography: A Review. Front Photonics 2022;3:1–16.

[43] Rivenson Y, Wu Y, Ozcan A. Deep learning in holography and coherent imaging. Light Sci Appl 2019;8:85.